\def\rn{\noindent\parshape 2 0truecm 8.5truecm 0.3truecm 8.2truecm}
\def\rn{}% NAME STYLE: Neumann, A. E. 
\def\nn#1 #2{#1, #2.}				% Name with 1 initial
\def\nnn#1 #2 #3{#1, #2. #3.}			% Name with 2 initials
\def\nnnn#1 #2 #3 #4{#1, #2. #3. #4.}		% Name with 3 initials
\def\nnnnn#1 #2 #3 #4 #5{#1, #2. #3. #4. #5.}	% Name with 4 initials
\def\dualand{ and\hbox{ }}				
\def\multiand{, and\hbox{ }}				
\def\rf#1;#2;#3;#4;#5 {\par\rn#1 #2, {\it #3}, {\bf #4}, #5\par}
\def\rg#1;#2;#3;#4;#5;#6 {\par\rn#1 #2, {\it #3}, {\bf #4}, #5 (``#6'') \par}
\def\rfbook#1;#2;#3;#4;#5 {{\frenchspacing\par\rn#1 #2, {\it #3} (#4: #5)\par}}
\def\rfproc#1;#2;#3;#4;#5;#6 {{\frenchspacing\par\rn#1 #2, in {\it #3}, ed. #4 (#5: #6)\par}}
\def\rfprocp#1;#2;#3;#4;#5;#6;#7 {{\frenchspacing\par\rn#1 #2, in {\it #3}, ed. #4 (#5: #6), p#7\par}}
\def\rfprep#1;#2;#3  {{\par\rn#1 #2, #3\par}}
\def\K{{\rm K}}
\def\mK{{\rm mK}}

\def\Jy{{\rm Jy}}
\def\mJy{{\rm mJy}}

\def\Mpc{{\rm Mpc}}
\def\MHz{{\rm MHz}}

\def\expec#1{\langle#1\rangle}

\def\etal{{\frenchspacing\it et al.}}
\def\ie{{\frenchspacing\it i.e.}}
\def\eg{{\frenchspacing\it e.g.}}
\def\etc{{\frenchspacing\it etc.}}

\def\beq#1{\begin{equation}\label{#1}}
\def\eeq{\end{equation}}
\def\beqa#1{\begin{eqnarray}\label{#1}}
\def\eeqa{\end{eqnarray}}
\def\eq#1{equation~(\ref{#1})}

\def\fig#1{Figure~\ref{#1}}
\def\Fig#1{Figure~\ref{#1}}

\def\sec#1{Section~\ref{#1}}

\def\subsec#1{Subsection~\ref{#1}}

\def\spose#1{\hbox to 0pt{#1\hss}}
\def\simlt{\mathrel{\spose{\lower 3pt\hbox{$\mathchar"218$}}
     \raise 2.0pt\hbox{$\mathchar"13C$}}}
\def\simgt{\mathrel{\spose{\lower 3pt\hbox{$\mathchar"218$}}
     \raise 2.0pt\hbox{$\mathchar"13E$}}}
\def\simpropto{\mathrel{\spose{\lower 3pt\hbox{$\mathchar"218$}}
     \raise 2.0pt\hbox{$\propto$}}}

\def\l{\ell}

\def\z{{\bf z}}

\def\N{{\bf N}}

\def\Ob{\Omega_{\rm b}}

\def\Ok{\Omega_{\rm k}}
\def\Ol{\Omega_\Lambda}
\def\Om{\Omega_{\rm m}}

\def\ns{n_s}

\def\l{\ell}

\def\dTb{\delta T_b}
\def\xe{x_e}
\def\dxe{\delta_{x_e}}

\documentclass{emulateapj}
\begin{document}
% \twocolumn[

%%%%%%%%%%%%%%%%%%%%%%%%%%%%%

%\tighten
%\eqsecnum
%\received{4 August 1988}
%\accepted{23 September 1988}
%\journalid{337}{15 January 1989}
%\articleid{11}{14}

%%%%%%%%%%%%%%%%%%%%%%%%%%%%%%%%%%%%%%%%%%%%%%%%%%%%
%%%%%%%%%%%%%%%%%%%%%%%%%%%%%%%%%%%%%%%%%%%%%%%%%%%%
\title{Twenty-one centimeter tomography with foregrounds}

\author{Xiaomin Wang}
\affil{Kavli Institute for Cosmological Physics, Univ. of Chicago, Chicago, IL 60637; xiaomin@cfcp.uchicago.edu}
\affil{Dept. of Physics, Univ. of Pennsylvania, Philadelphia, PA 19104}
\author{Max Tegmark}
\affil{Dept. of Physics, MIT, 70 Vassar Street, Rm. 37-626B, Cambridge, MA 02139}
\affil{Dept. of Physics, Univ. of Pennsylvania, Philadelphia, PA 19104}
%\author{Mario G. Santos \and Lloyd Knox}
\author{Mario G. Santos}
\affil{CENTRA, Instituto Superior Tecnico, Universidade Tecnica de Lisboa,
Portugal \\
Dept. of Physics, Univ. of California at Davis, Davis, CA 95616}
\author{Lloyd Knox}
\affil{Dept. of Physics, Univ. of California at Davis, Davis, CA 95616}

%%%%%%%%%%%%%%%%%%%%%%%%%%%%%%%%%%%%%%%%%%%%%%%%%%%%%%%%%%%%
%%%%%%%%%%%%%%%%%%%%%%%%%%%%%%%%%%%%%%%%%%%%%%%%%%%%%%%%%%%%

\begin{abstract} 
Twenty-one centimeter tomography is emerging as a powerful tool to 
explore the reionization epoch and cosmological parameters,
but it will only be as good as our ability to accurately model and remove astrophysical foreground contamination.
Previous treatments of this problem have focused on the angular structure of
the signal and foregrounds and what can be achieved with limited spectral resolution 
(channel widths in the 1 MHz range).
In this paper we introduce and evaluate a ``blind'' method to extract the multifrequency 21cm signal 
by taking advantage of the smooth frequency structure of the Galactic and extragalactic foregrounds. 
We find that 21 cm tomography is typically limited by foregrounds on scales $k\ll 1h/$Mpc
and limited by noise on scales $k\gg 1h/$Mpc, provided that the experimental channel width can be made
substantially smaller than 0.1 MHz.
Our results show that this approach is quite promising even for scenarios with rather extreme 
contamination from point sources and diffuse Galactic emission,
which bodes well for 
upcoming experiments such as LOFAR, MWA, PAST, and SKA.
\end{abstract}
% ]  % Must end \twocolumn command here, or disaster occurs, for bizarre reasons.

%%%%%%%%%%%%%%%%%%%%%%%%%%%%%%%%%%%%%%%%%%%%%%
%%%%%%%%%%%%%%%%%%%%%%%%%%%%%%%%%%%%%%%%%%%%%%

\section{Introduction}

21 cm tomography is one of the most promising cosmological probes, with the potential to complement 
and perhaps ultimently eclipse the cosmological parameter constraints from 
the cosmic microwave background \citep{Bowman05,McQuinn05}.
It is also a unique probe of 
the epoch of reionization, which is now one of the least understood aspects 
of modern cosmology.
%We are still largely in the dark about 
%the epoch of reionization, which is now one of the least understood aspects 
%of modern cosmology.
%On one hand, 
%Gunn-Peterson constraints \citep{GunnPeterson} from quasars require the Universe to be highly
%ionized by $z\sim 6$  
%%(\eg, \citep{Becker01,Fan02}).
%\citep[\eg,][]{Becker01,Fan02}.
%On the other hand, WMAP data \citep{Kogut03,Spergel03} suggest a high
%reionization optical depth $\tau\sim 0.17$, which could be explained by the Universe being reionized at redshift $z\sim 17$,
%so the actual reionization history of the Universe may have been more complex and interesting 
%than a complete reionization at a single epoch.
There are various techniques to explore the epoch of
reionization at $5<z<20$. Apart from the CMB \citep{HHKK03,Knox03,Kogut03,Santos03}, 
radio astronomical measurement of 21cm radiation from neutral hydrogen has been shown theoretically to be 
a powerful tool to study this period \citep{MMR,Tozzi99}. 
Lots of work has been done in recent years on various theoretical and experimental aspects of 21cm radiation 
 \citep[\eg,][] 
{BarkanaLoeb04a,Carilli02,Carilli04a,CiardiMadau03,DiMatteo02,DiMatteo04,Furlanetto03,FZH04,GnedinShaver03,Iliev02a,
Iliev02b,Loeb03,McQuinn05,Morales04,OhMack03,Pen04,Santos04,Shaver99,
WyitheLoeb04a,WyitheLoeb04,Matias03}.

However, this 21cm tomography technique 
will only be as good as our ability to accurately model and remove astrophysical foreground contamination,
since the high redshift signal one is looking for is quite small and can be easily swamped  
by foreground emission from our galaxy or others. 
With much effort going into upcoming experiments such as the 
Mileura Wide-Field Array (MWA)\footnote{http://web.haystack.mit.edu/MWA/MWA.html}, 
% \citep{mileura},
LOFAR\footnote{http://www.lofar.org} \citep{Rottgering03}, 
PAST\footnote{http://astrophysics.phys.cmu.edu/~jbp},
% \citep{past}
and SKA\footnote{http://www.skatelescope.org},
% \citep{ska}, 
aimed at gathering redshifted 21cm signal from the sky and probe the epoch of
reionization, it is therefore timely to study the foreground problem in detail. 

Although 21cm foregrounds have been discussed in some previous papers, 
\citep[\eg,][]{DiMatteo02,DiMatteo04,MoralesHewitt03,OhMack03,Santos04,Matias03}, the questions on how to remove foregrounds
and noise from observations of 21cm signal, how well it can be done, and how reliable it is,
are still wide open. 
Previous papers have focused on the angular power spectrum of the signal, usually assuming a rather limited spectral resolution \citep{DiMatteo02,DiMatteo04,OhMack03,Santos04,Matias03}. In this paper, we develop a method to remove the foregrounds along the line of sight, 
taking advantage of the fact that most astrophysical contaminants have much smoother frequency spectra than the 
cosmological signal one is looking for. The two approaches are complementary, and we will argue that they
are best used in combination: our technique can be used both to identify point sources and other 
highly contaminated angular regions to be discarded, and to clean out residual contamination 
from those angular regions that are not discarded.
This multifrequency approach is more powerful here than for typical CMB applications
\citep{Bennett03,Max98foreground,TOH03}, because of the potentially much better 
spectral resolution, and the dramatically oscillating 21cm signal compared with
smooth foregrounds along frequency direction.

In this paper, we describe the method for removing foregrounds 
% and detector noise 
in frequency space, show
examples of using this method in different scenarios, and discuss its promising applications for future
experiments. 
In \sec{briefmethodsec}, we introduce the reionization model we use throughout the text, then give a brief overview of 
21cm emission/absorption and computational formalism, 
on how we calculate the 21cm angular power spectrum in $\l$-space, 
projected 1D and 2D power spectra in $k$-space, and the simulated 1D frequency spectrum in real space. 
% Appendix gives technical details of this and our linearization procedure.
In \sec{cleanmethodsec}, we describe our foregrounds removing strategy, we also show the foregrounds model we use in our calculations. 
In \sec{cleanexamplesec}  we give several
applications of our method under different assumptions about foregrounds and noise.  
% for patch of sky with  
% only one kind of foreground, and one with everything including strong point source contamination,\etc\
% including both a patch of sky with negligible point source foreground and one with strong point source contamination, \etc\     
We summarize our results in \sec{discsec}.

\section{Reionization model and formalism}
\label{briefmethodsec}

The reionization model we 
use throughout this paper is from \citep{HaimanHolder03, Santos03}, shown as the solid curve in 
Fig 3 in \citep{Santos03}. 
% \fig{xeFig}. 
Although the most
recent results from WMAP \citep{spergel06} favors a lower optical depth, 
the results of this paper are rather insensitive to the detailed choice of reionization model 
and associated
assumptions, since we are focused on foregrounds rather than the cosmological 21 cm signal.
For more information 
about various reionization models, see, for example, \citep{HaimanHolder03,HHKK03,Santos03}. 

% \begin{figure}[h] 
% %\vskip-1.0cm
% \centerline{\epsfxsize=8.5cm\epsffile{xe.eps}}
% \bigskip
% %\vskip-1.0cm
% \caption{\label{xeFig}\footnotesize%
% Ionization fraction $x_e$ for three different fiducial models with $\tau=0.17$.
% The solid curve is the model we use. 
% From Haiman \& Holder \protect\citep{HaimanHolder03} 
% % (see also Santos et al.\protect\citep{Santos03}).
% \protect\citep[see also][]{Santos03}.
% }
% \end{figure}

Below we give a brief overview of the 21cm emission power spectrum and our calculational method. 
The detailed information on 21cm radiation
(emission/absorption) phenomena can be found in various literatures, \eg, \citep{MMR,Santos04,Shaver99,Tozzi99,Matias03}.

\subsection{3D power spectrum}

% In case of 21cm emission, 
The differential antenna temperature observed at Earth between the neutral
hydrogen patch and the CMB can be approximated as \cite{Shaver99,Tozzi99}
\beq{dtbpEq}
\dTb\approx 0.016\K {1\over h}(1+\delta)(1-x){{\Omega_b h^2}\over 0.02}
         \left[{{1+z}\over 10}{0.3\over \Om}\right]^{1/2}, 
\eeq
where $\delta=\rho/\bar{\rho} -1$ is the fluctuation of the density field.  
We write the ionization fraction $x$ as a sum of two terms \citep{Santos03} 
\beq{xeeq}
x=\xe (1+\dxe),
\eeq
where $\xe$ is average ionization fraction and $\dxe$ is the fluctuation of the ionization fraction
across the sky. Thus the ionization fraction $x$ is not only a function of redshift, but also dependent
on its position in the sky. 

Assuming $\delta \ll 1$ and $\dxe \ll 1$  
 and neglecting all second and higher order terms, 
we obtain the 3D power spectrum for 21cm emission,  
\beqa{Pk3deq3}
P_{3D}(k,z)=(0.016\K)^2 {1\over h^2}\left({{\Ob h^2}\over 0.02}\right)^2 {{1+z}\over 10}{0.3\over \Om}\nonumber \\
\times (1-2\xe+\xe^2+b^2 e^{-k^2 R^2}\xe^2)\ {P_{matter}(k)}.
\eeqa
% \beqa{Pk3deq2}
% &&P_{3D}(k,z)\delta^3_D(k-k')=\expec{(\dTb - \expec{\dTb})(\dTb' - \expec{\dTb'})}    \nonumber \\  
% &&=(0.016\K)^2 {1\over h^2}\left({{\Ob h^2}\over 0.02}\right)^2 {{1+z}\over 10}{0.3\over \Om}\nonumber \\
% &&\times [(1-\xe-\xe'+\xe\xe')\expec{\delta\delta'}+\xe\xe'\expec{\dxe\dxe'}],
% \eeqa
where $P_{matter}(k)$ is the matter power spectrum. 
% $\expec{\delta\delta'}=P_{\delta}(k)\delta^3_D(k-k')=P_{matter}(k)\delta^3_D(k-k')$,  $P_{matter}(k)$ is the matter power spectrum. 
And the power spectrum for the ionized fraction $P_{\dxe}(k)$  
% for which $\expec{\dxe\dxe'}=P_{\dxe}(k)\delta^3_D(k-k')$, 
is defined as in \citep{Santos03}. 
\beq{Pioneq}
P_{\dxe}(k)=b^2 P_{matter}(k) e^{-k^2 R^2},
\eeq
where $b$ is the mean bias weighted by the different halo properties
. The mean radius $R$ of the ionized patches in HII regions is modeled as
\beq{radiuseq}
R=\left ({1\over {1-\xe}}\right )^{1/3} R_p,
\eeq
where $R_p$ is the comoving size of the fundamental patch, $R_p\sim 100$Kpc \citep{Santos03}.

% Substituting \eq{Pioneq} into \eq{Pk3deq2}, the 3D power spectrum for the observable signal (fluctuation in the difference
% between spin temperature and CMB temperature across the sky) can finally be written as 
% \beqa{Pk3deq3}
% P_{3D}(k,z)=(0.016\K)^2 {1\over h^2}\left({{\Ob h^2}\over 0.02}\right)^2 {{1+z}\over 10}{0.3\over \Om}\nonumber \\
% \times (1-2\xe+\xe^2+b^2 e^{-k^2 R^2}\xe^2)\ {P_{matter}(k)}.
% \eeqa

We assume a cosmological concordance model 
\citep{sdsslyaf,Spergel03,sdsspars}
($\Ok=0$, $\Ol=0.71$, $\Ob=0.047$, $h=0.72$, $\ns=0.99$, $\sigma_8=0.9$) throughout our calculations.

\subsection{Projected 1D power spectra} 

The 21 cm signal changes with redshift for two separate reasons, 
one slow and one fast:
\begin{enumerate}
\item The average properties of the Universe ($x$, $T_k$, $T_s$, \etc) evolve on a
timescale $\Delta\z \sim 1$.
% (Figure~\ref{xeFig}).
\item The local properties of the Universe change on much smaller scales $\Delta z$ corresponding 
to the sizes and separations between ionized regions.
\end{enumerate}
Across a very small redshift range $z_0-\Delta z<z<z_0+\Delta z$ where $\Delta z < 1$, 
we make the approximation of ignoring the former and including only the latter,
approximating parameters like $x$, $T_k$ and $T_s$ by their values at $z_0$.
This enables us to linearizes the relations between 
frequency $\nu$, redshift $z$ and comoving radial distance $D$.

Making the above-mentioned approximation and ignoring redshift space distortions, 
the 21 cm signal near a given $z_0$ has an isotropic 3D power spectrum $P_{3D}$ that we can project 
into 1D (radial) power spectra $P_{1D}(k,z_0)$ 
\citep{Hui99,Peacockbook}:
\beq{Pk1deq}
P_{1D}(k,z\approx z_0)={1\over {2\pi}}\int_k^{\infty}P_{3D}(k',z\approx z_0)k'dk',
\eeq

% \beq{Pk2deq}
% P_{2D}(k,z\approx z_0)={1\over {2\pi}}\int_k^{\infty}P_{3D}(k',z\approx z_0){{k'}\over{\sqrt{{k'}^2-k^2}}}dk'
% \eeq
% \clearpage
\begin{figure}[ht] 
%\vskip-1.0cm
%\centerline{\epsfxsize=8.5cm\epsffile{fxnu_9_dz01N50.ps}}
\centerline{\epsfxsize=8.5cm\epsffile{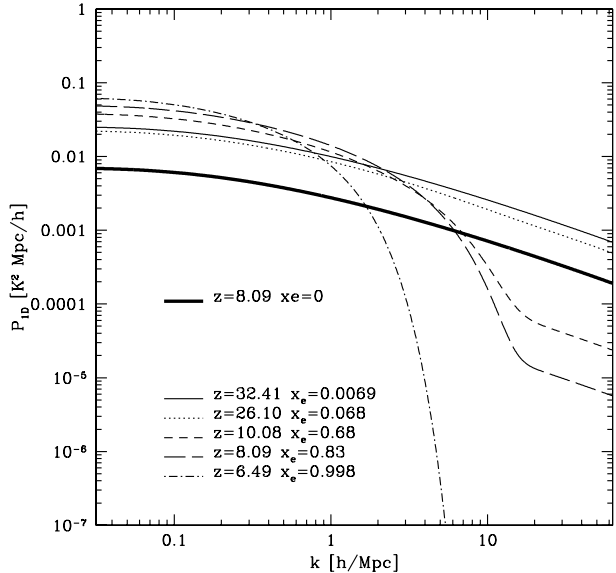 }}
% \centerline{\includegraphics[width=8.5cm]{f1.ps}}
\bigskip
%\vskip-1.0cm
\caption{\label{p1dFig}%
Line-of-sight 1D 21cm power spectra at different redshifts and ionization fractions. 
The thick solid line is power
spectrum for neutral IGM at redshift 8.09. 
}
\end{figure}
% \clearpage
\Fig{p1dFig} shows the line of sight 1D 21cm emission power spectra for the fiducial reionization model 
at different reionization epochs. For comparison, we also plot the power spectrum for neutral
medium at $x=8.09$ (thick solid line).

\subsection{Simulated signal in real space from 1D power spectrum}
\label{simsigsec}

We generate and analyze our simulations with fast Fourier transforms.
The simulated signal in real space in the region $0<x<L$ is 
\beq{FT1deq}
f(x)={1\over \sqrt{N}}\sum_{q=0}^{N-1}\left[A_q\cos\left ({2\pi x\over L}q\right)+B_q\sin\left({2\pi x\over L}q\right)\right]
\eeq
where $A_q$ and $B_q$ are Gaussian random variables with zero mean and standard deviations
$\Delta A_q = \Delta B_q =\sqrt{P_{1D}(k,z_0)/2}=\sqrt{P_{1D}(2\pi q/L,z_0)/2}$. 
N is chosen to be a large enough integer that all information from $P_{1D}(k)$ is included in the summation 
($k_{max}=2\pi N/L$) . 
The box size L should be small enough so that the range of L satisfies $\Delta z\ll 1$. 
% According to \eq{Dlineq}, this is equivalent to  
% \beq{Llimiteq}
% L\leq D'(z_0)|z-z_0|=D'(z_0)\Delta z
% \eeq
% \clearpage
\begin{figure}[tb] 
%\vskip-1.0cm
%\centerline{\epsfxsize=8.5cm\epsffile{fxnu_9_dz01N50.ps}}
\centerline{\epsfxsize=8.5cm\epsffile{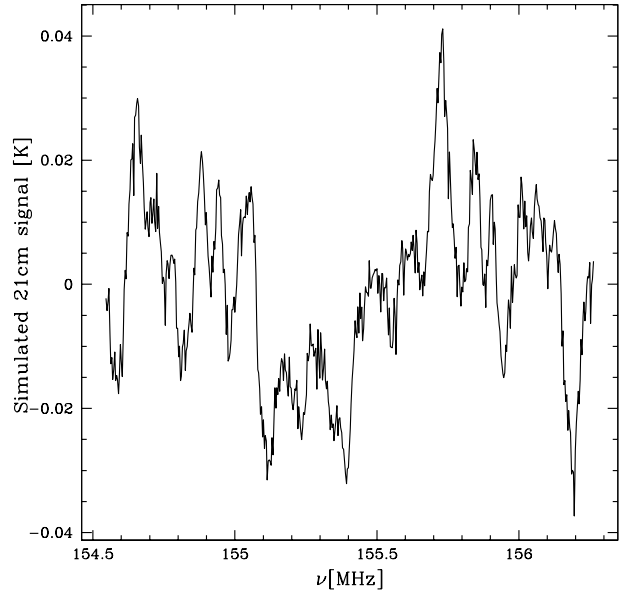 }}
% \centerline{\includegraphics[width=8.5cm]{f2.ps}}
\bigskip
%\vskip-1.0cm
\caption{\label{sigFig}%
Simulated 21cm signal in frequency space corresponding to the $z\approx 8.09$ and $x_e\approx 0.83$.
%$x_e\approx 0.827$.
}
\end{figure}
% \clearpage
\Fig{sigFig} shows our simulated 21cm signal versus frequency around 155 MHz, corresponding to 
an epoch around redshift $z_0=8.09$.

If we plotted the {\it observed} signal in the relevant frequency range,
the expected contribution from foregrounds would lie far above the cosmological signal shown in.
The key to doing 21cm cosmology is therefore removing foregrounds using multi-frequency information, 
as emphasized by, \eg, \citep{Matias03}, and we now turn to this subject.

\section{Method for foreground removal in frequency space}
\label{cleanmethodsec}

Because of its small frequency cross-correlations, the 21cm signal is oscillating dramatically along the frequency direction. The foregrounds, on the other hand, are generally quite smooth over the short frequency range we consider.
This slowly varying nature of the foregrounds compared to 
the signal is a great advantage when removing it \citep{GnedinShaver03,McQuinn05,
MoralesBowmanHewitt05}, and it is the main reason
that our foreground removal method works so well.

Our method described here is insensitive to the reionization model 
and the redshift range we choose,  since we are focused on foregrounds rather than the cosmological 21 cm signal.

\subsection{Foreground removal method}

Our basic approach is to subtract foregrounds separately in each angular direction in the sky, by fitting 
their total intensity dependence of frequency by a log-log polynomial.
Note that since we are fitting the total foreground spectrum separately pixel by pixel 
(fitting not only for the amplitude but also for the spectral index and the running of the spectral index),  
we are unaffected by the possible complication of huge variations of the foreground spectral index across the sky. 
(If the foregrounds would lack both frequency coherence and spatial coherence, \ie, fluctuate randomly with both frequency and position,
then we would be unable to identify and remove them and could merely average them down like we do with detector noise.)

There are two separate steps in our analysis: 
\begin{enumerate}
\item Simulation 
\item Cleaning
\end{enumerate}
We treat them as completely independent. In other words, our cleaning algorithm is {\it blind}, containing 
no information about the foreground and noise model used in the simulation step. 
It is entirely specified by the single
integer $m$ giving the order of the log-log fitting polynomial.

In the simulation step, we simulate for each pixel the total observed
frequency spectrum $y_i$ at $n$ different log-frequencies $x_i = \lg(\nu_i)$, $i=1,...,n$.  
This simulated total signal $y_i=\lg{(I_{21cm}^i+I_{syn}^i+I_{ff}^i+I_{ps}^i+I_{det}^i)}$, 
includes 21cm signal $I_{21cm}^i$, synchrotron emission foreground $I_{syn}^i$, free-free emission
foreground $I_{ff}^i$, point source foreground $I_{ps}^i$ and detector noise $I_{det}^i$. 
% \beq{ytotaleq}
% % y_i=I_{21cm}^i+I_{syn}^i+I_{ff}^i+I_{ps}^i+I_{det}^i
% y_i=\lg{(I_{21cm}^i+I_{syn}^i+I_{ff}^i+I_{ps}^i+I_{det}^i)},
% % y_i = \lg{(I_{21cm}+I_{synchrotron}+I_{detector}+\cdots )}. 
% \eeq
We test a variety of different assumptions for foregrounds and noise in this step.

Then we group the $y_i$'s into an $n$-dimensional vector  ${\bf y}$,
% \beq{ymatrix}
% {\bf y}=\left (\begin{array}{c}
%          y_1  \\
%          y_2  \\
% 	            ...        \\
%          y_n
%   \end{array}\right )
% \eeq
and group the $x_i$'s and their powers into 
an $n\times m$ matrix ${\bf X}$. 
% \beq{xmatrix}
% {\bf X}=\left (\begin{array}{cccc}
% 	 x_1^{m-1} & ... & x_1 & 1 \\
% 	 x_2^{m-1} & ... & x_2 & 1 \\
% 		  ...	     \\
% 	 x_n^{m-1} & ... & x_n & 1 
%   \end{array}\right ),
% \eeq
So that the data can be modeled as
\beq{fiteq}
{\bf y}={\bf X}{\bf a} + {\bf n},
\eeq
where the $m$-dimensional parameter vector ${\bf a}$
% \beq{amatrix}
% {\bf a}=\left (\begin{array}{c}
%          a_1  \\
%          a_2  \\
% 	            ...        \\
%          a_m
%   \end{array}\right )
% \eeq
parametrizes the foreground contributions. It is what we need to find out in the cleaning step. 
In \eq{fiteq}, ${\bf n}$ is the part left in the total signal that can not be fitted by the parameters in {\bf a}, 
including the contribution from signal, detector noise and residual foregrounds.
% For a linear fit, $m=2$ so that $y_i=a_1x_i+a_2$.

In all our calculations throughout this paper, we take this fitting polynomial to be quadratic, \ie, 
fit the total foregrounds as a single running power law. 
Equivalently we fit the log intensity of the foreground as 
$\lg I=a_3+a_2\lg\nu+a_1(\lg\nu)^2$. 
That is to say, the number of fitting parameter in our computation is always $m=3$. 
We found that this improved noticeably over $m=2$, whereas increasing
to $m=4$ gave essentially no further improvement.

In the cleaning step, 
% on a pixel by pixel basis, we fit to the total
% signal ${\bf y}$, obtain a best fit foreground model with parameters ${\bf a}$, subtract it from the total signal, and recover a cleaned
% signal.
we find our best fit parameter vector {\bf a} by minimizing $\chi^2=({\bf y}-{\bf X}{\bf a})^t{\bf N}^{-1}({\bf y}-{\bf X}{\bf a})$,
obtaining \citep{mapmaking}
\beq{bestpfiteq}
{\bf a}=[{\bf X}^t{\bf N}^{-1}{\bf X}]^{-1}{\bf X}^t{\bf N}^{-1}{\bf y},
\eeq
where ${\bf N}$ is the covariance matrix of the contributions from the detector noise and 21cm signal. 
We then subtract the total fitted foreground contribution 
${\bf X}{\bf a}$ from the simulated measurement vector, thus obtaining what we will refer to as the cleaned signal {\bf y}. 

Although this cleaning technique is only optimal if $\N$ is known and the contributions from noise 
and 21cm signals are Gaussian \citep{mapmaking}, we use \eq{bestpfiteq} anyway and quantify the residual noise using 
our simulations. 
Since \eq{bestpfiteq} minimizes the rms residual even in the presence of non-Gaussianity \citep{mapmaking}, it is a robust general-purpose 
fit that does not require detailed foreground or signal modeling.
We simply set ${\bf N}={\bf I}$, the identity matrix,
which will be essentially optimal if white detector noise dominates. 
If desired, this can be further improved by modeling the foreground power spectrum found in real data and iterating.

Since the cleaning step uses a single polynomial in log-log, it cannot fit exactly a simulation including 
detector noise
or more than one foreground component (since adding the exponential of two polynomials does not give the exponential of a polynomial).
We will see that this simple cleaning algorithm is nonetheless very successful, able to fit any of our foreground 
models well over the limited frequency range that is relevant.

\subsection{Foreground and detector noise models}

The foregrounds we consider in this paper include Galactic synchrotron emission, Galactic free-free
(thermal) emission, and extragalactic point sources. For more information on
models of foregrounds in the 100 MHz range, please see, \eg, 
\citep{DiMatteo02,DiMatteo04,Haslam82,Haverkorn03,MoralesHewitt03,OhMack03,Platania03,Santos04,Shaver99}.
As emphasized above, the foreground models we describe here are used only in our simulation step,
not for the cleaning process.

\subsubsection{Galactic synchrotron radiation}

For Galactic synchrotron emission, which probably causes most contamination of all foregrounds
(perhaps of order 70\% at 150MHz \citep{Platania03,Shaver99}), 
we assume its intensity to be a running power law in frequency
\beq{synrunningeq}
I_{syn}=A_{syn}\left(\nu\over{\nu_*}\right)^{-\alpha_{syn}-\Delta\alpha_{syn}\lg{{\nu}\over{\nu_*}}}, 
\eeq
with a spectral index $\alpha_{syn}=2.8$ \citep{Max99foreground} and
a spectral index ``running'' $\Delta\alpha_{syn}=0.1$ \citep{Haverkorn03,Platania03,Shaver99,Max99foreground}.
Here $\nu_*\equiv 150\MHz$. 
We assume the amplitude of the synchrotron foreground to be $A_{syn}=335.4\K$, an extrapolation from
 \citep{Haverkorn03,Max99foreground}. We also explore other normalizations $A_{syn}$ orders of magnitude higher than the
value we define here in our calculations. Similarly we try other values of spectral index and spectral
running index in the calculation to test the robustness of our method. We discuss the details
in \sec{cleanexamplesec}.

\subsubsection{Galactic free-free emission}

We model the Galactic free-free emission (which might contribute a contamination of order 
1\% contamination at 150MHz \citep{Shaver99}) as a
running power law as well \citep{Haverkorn03,Platania03,Max99foreground}, 
\beq{ffrunningeq}
I_{ff}=A_{ff}\left(\nu\over{\nu_*}\right)^{-\alpha_{ff}-\Delta\alpha_{ff}\lg{{\nu}\over{\nu_*}}},
\eeq
where $\alpha_{ff}=2.15$, $\Delta\alpha_{ff}=0.01$, 
and $A_{ff}=33.5\K$, extrapolated from \citep{Haverkorn03,Max99foreground}.

\subsubsection{Extragalactic point sources}

Point sources have been estimated to cause about 30\% of the contamination at 150MHz \citep{Shaver99} and are typically
less smooth in frequency than the Galactic foregrounds. When looking in a given direction, we are observing the same
point sources as we change frequency, so there are not small-scale fluctuations
in the same sense as when we change observing directions\footnote{There is, however, 
the subtle effect of off-beam point sources dimming towards higher frequencies because the beam gets narrower 
 \citep{OhMack03,Matias03}.
For the narrow frequency intervals $\Delta\ln\nu$ that we are considering, this effect will be around the
percent level for individual sources, in the same ballpark as the intensity change due to the frequency dependence of the 
emission mechanism. This means that it will not imprint sharp spectral features in
the total foreground emission, and should be well fit by our blind cleaning algorithm. We have not included
this complication in the present analysis --- it would be worth incorporating it in a more detailed foreground analysis,
particularly in one including explicit modeling of the sky pixelization.  
%(the subtle effect of off-beam point sources dimming towards higher frequencies because the beam gets narrower
%should be a rather small correction considering the small frequency intervals $\Delta\ln\nu$ we are considering).
}.
A serious complication compared to the Galactic synchrotron and free-free cases is that
when we observe many point sources in a pixel, they can each have quite different spectral indices,
possibly making their combined intensity
a quite complicated function of frequency. 

One approach would be to model this complicated function as a 
running power law over the narrow frequency range involved, just as we did for 
the synchrotron and
free-free foregrounds:
\beq{pseq}
I_{ps}=A_{ps}\left ({S_{cut}\over\rm{mJy}}\right )^\beta \left ({{\nu}\over{\nu_*}}\right )^{-\alpha_{ps}-\Delta\alpha_{ps}\lg{{\nu}\over{\nu_*}}},
\eeq
where $\alpha_{ps}=2.81$, $\Delta\alpha_{ps}=0.25$ and $\beta=0.125$ \citep{Max99foreground}. 

However, we wish to be as conservative as possible in this paper, and therefore adopt a more complicated point source model
in our simulations.
Instead, we therefore simulate a large number of random point sources $i=1,...$ in the pixel that we are considering 
and sum their intensity contributions in Kelvin: 
\beq{psgaus2eq}
I_{ps} = \left({dB\over dT}\right )^{-1}  \Omega_{\rm sky}^{-1}\sum_i S^*_i \left(150\>\MHz\over\nu\right)^{\alpha_i}-\expec{I_{ps}},
\eeq
where  
\beq{avepsgaus2eq}
\expec{I_{ps}} = \left ({dB\over dT}\right )^{-1}\int_0^{S_{cut}} {S{{dN}\over{dS}}\ dS} \int { \left ({150\over\nu}\right )^\alpha f(\alpha)\ d\alpha} 
\eeq
is the average value of point source foreground intensity.
Conversion factor 
$dB/dT=6.9\times 10^5 \mJy/\K$. The assumed sky area per pixel is approximately 
% $\Omega_{\rm sky}=4\times 10^{-8}\rm{sr}$. 
% $\Omega_{\rm sky}=10^{-6}\rm{sr}$. 
$\Omega_{\rm sky}=12\ \rm{arcmin^2}$. 
In \eq{psgaus2eq}, 
$S^*_i$ is the flux of the $i^{\rm th}$ point source at 150 MHz. It is generated randomly from
the source count distribution $dN/dS=4(S/1\Jy)^{-1.75}$ \citep{DiMatteo04}, truncated at a maximum flux $S_{\rm
max}=S_{cut}=0.1\mJy$, above which we
assume that point sources can be detected and their pixels discarded. In other words, when we talk about the contamination 
from point sources below, we refer only to the contribution from unresolved point sources. 
To avoid having to generate infinitely many point sources, we also truncated the distribution at a
minimum flux $S_{\rm min}=10^{-3}\mJy$, since we find that the total flux contribution has converged by then.
We generate $\alpha_i$, the spectral index of the $i^{\rm th}$ point source, randomly from
the Gaussian distribution  
\beq{gauseq}
f(\alpha)={1\over {\sqrt{(2\pi)}\sigma_\alpha}} \exp{\left
[-{{(\alpha-\alpha_0)^2}\over{2\sigma_\alpha^2}}\right ]}, 
\eeq
with spectral index $\alpha$ in the range of $[\alpha_0-\Delta\alpha,\alpha_0+\Delta\alpha]$, where
 $\Delta\alpha=5\sigma_\alpha$. To be conservative, we allow the spectral index vary
 in a fairly large region, $\sigma_\alpha=10$,  through our calcuations.

\subsubsection{Detector noise}

We treat detector noise as white noise. 
In the Rayleigh-Jeans limit, the 
rms detector noise
% detector noise variance 
in a pixel can be approximated as 
\beq{pixelnoise}
\sigma_T={\lambda^2\over {2k_B}}B={\lambda^2\over {2k_B}}{S\over A}
\eeq
where $k_B$ is Boltzmann constant, $\lambda$ is redshifted wavelength of 21cm emission. The specific
brightness $B$ is related to point source sensitivity $S$ by dividing it with pixel area $A$. 

At redshift 8.47, $\nu=150\MHz$, $\lambda=2m$, with LOFAR virtual core
configuration\footnote{http://www.lofar.org}, for a
5.2\arcmin\ pixel with $4\MHz$ bandpass and one hour integration, sensitivity $S$ is approximately 0.17\mJy, from \eq{pixelnoise}, we get 
\beq{lofarnoise}
\sigma_T^{\rm{LOFAR}}=108\mK \left ({{4\MHz}\over{\Delta\nu}}\right )^{0.5} \left ({{1\ \rm{hour}}\over{t}}\right )^{0.5}
\eeq
where $\Delta\nu$ is channel width and $t$ is total integration time. 
Similarly for MWA experiment\footnote{http://web.haystack.mit.edu/MWA/MWA.html}, a 4.6\arcmin\ pixel with $32\MHz$ bandpass and one hour integration,
point source sensitivity $S=0.27\mJy$, so we get MWA detector noise 
\beq{mwanoise}
\sigma_T^{\rm{MWA}}=218\mK \left ({{32\MHz}\over{\Delta\nu}}\right )^{0.5} \left ({{1\ \rm{hour}}\over{t}}\right )^{0.5}
\eeq

We should mention that although at 4MHz bandwidth, the sensitivity for MWA is worse than LOFAR, MWA
has a larger bandpass and field of view. This larger field of view leads to vastly more
pixels, which is an advantage for foregrounds removal, as we will see in later
sections. The detector thermal noise is only one of the many concerns
in the experiment, such as calibration, systematics, \etc\ Thus it should not be considered as the
only criterion to judge an experiment. 

The 1-dimensional power spectrum of the detector noise can then be written as
\beq{noisepowereq}
P_{det}=2\pi\sigma_{T}^2.
\eeq
In our simulation, we consider two scenarios. One scenario assumes a fiducial future 
experiment with Gaussian random detector noise down to $\sigma_T=1 \mK$ level. The other 
scenario assumes a currently achievable detector noise level of $\sim 200\mK$. 
This is based on \eq{lofarnoise} and
\eq{mwanoise} for the LOFAR and MWA experiments, assuming 1000 hours of integration
time and $4\rm{kHz}\sim 8\rm{kHz}$ frequency resolution, respectively.

\section{Results}
\label{cleanexamplesec}

% In this section, we present the results of applying the foreground removal 
% method described in \sec{cleanmethodsec} to our simulations.

As we showed previously in \fig{sigFig}, the signal wiggles rapidly with frequency. This is the key 
advantage of removing foregrounds in
frequency space, since foregrounds are typically relatively smooth functions of frequency.

% As mentioned above, 
We simulate
the 21 cm signal as a Gaussian random
field, although in reality, the signal is of course highly non-Gaussian. We make this Gaussianity approximation for
simplicity, since the key quantity that we are interested in (the power spectrum of the residual noise and
foregrounds) depends mainly on the power spectrum of the signal, foregrounds and noise, not on whether the statistics
are Gaussian or not.

%\subsection{Baseline example 1 --- ideal scenario (noise$\ll$signal)}
\subsection{Baseline example 1 --- long term potential (noise$\ll$signal)}
\label{lownoiseresult}

\begin{figure}[ht] 
%\vskip-1.0cm
% \centerline{\epsfxsize=8.5cm\epsffile{fxnu_allfg.ps}}
\centerline{\epsfxsize=8.5cm\epsffile{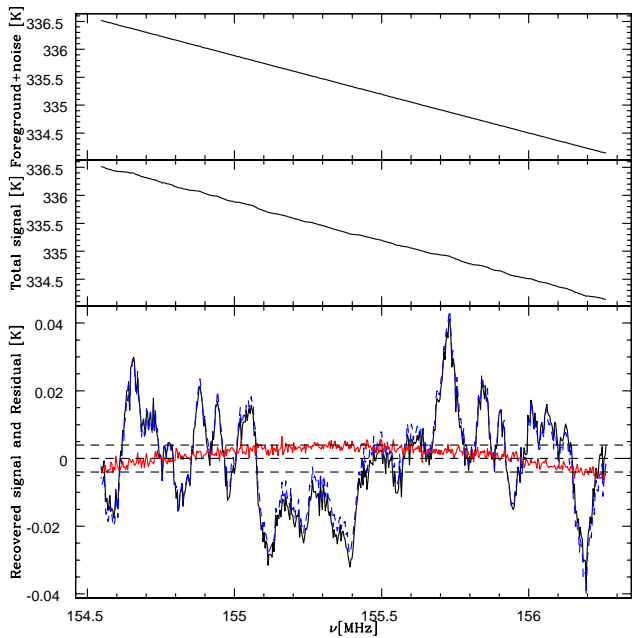}}
% \centerline{\includegraphics[width=8.5cm]{f3.ps}}
\bigskip
%\vskip-1.0cm
\caption{\label{midsyn_midff_ps_middet_Fig}%
Spectrum in a single pixel before and after foreground cleaning.
The top panel shows total contaminant signal, consisting of 
synchrotron radiation ($A_{syn}=335.4\K$, $\alpha_{syn} =2.8$, $\Delta\alpha_{syn} =0.1$), 
free-free emission foreground ($A_{ff}=33.5\K$, $\alpha_{ff} =2.15$, $\Delta\alpha_{ff} =0.01$),  
extragalactic point sources ($\sigma_\alpha=10$)
and detector noise ($\sigma=1\mK$).
The middle panel has the cosmological 21cm signal added. 
The bottom panel shows the recovered 21cm signal (blue curve) compared with the true simulated signal (black curve) 
and the residual (recovered minus simulated 21cm signal; red curve). 
The three horizontal black dashed lines correspond to -0.004K, 0K, and 0.004K, respectively. 
(Note the different vertical axis limits.) The small-scale wiggles in the residual are detector noise,
whereas the smoothed parabola-shaped component of the residual is the error in the foreground fitting.
}
\end{figure}

The results for the baseline example  with noise much smaller than the signal are shown in 
\fig{midsyn_midff_ps_middet_Fig}. The top panel shows the total contaminant in a pixel
including Galactic
synchrotron radiation, Galactic free-free emission, extragalactic point sources
and detector noise with $\sigma=1\mK$, which is the fiducial value for a future generation 
experiment. 
The foregrounds are modeled as in the previous section with parameters (given in the figure caption)
corresponding to rather pessimistic assumption about the foreground properties.

\begin{figure}[ht] 
%\vskip-1.0cm
\centerline{\epsfxsize=8.5cm\epsffile{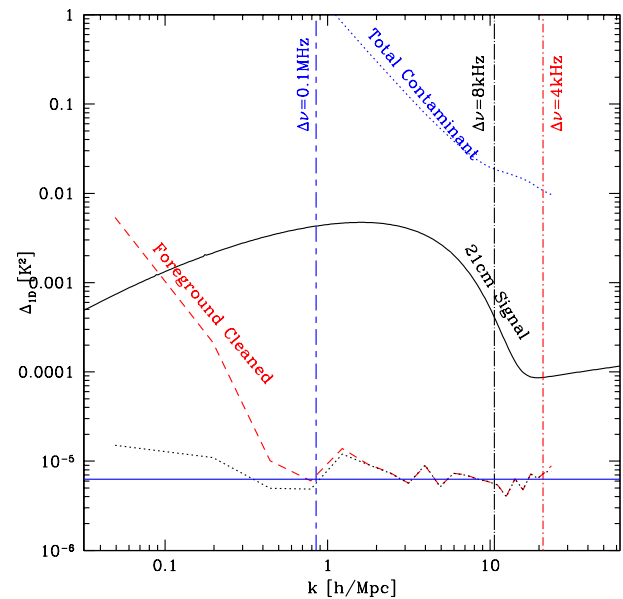}}
%\centerline{\includegraphics[width=8.5cm]{f4.ps}}
%\centerline{\epsfxsize=8.5cm\epsffile{p1d_5p8.ps}}
\bigskip
%\vskip-1.0cm
% \caption{\label{p1d_all_extreme_Fig}\footnotesize%
\caption{\label{p1d_midsyn_midff_Fig}%
1D power spectra of 21cm signal (solid black curve), total contaminant from \fig{midsyn_midff_ps_middet_Fig} 
(dotted blue curve), residual contaminant after foreground cleaning (dashed red curve)
and detector noise alone (dotted black curve).
The horizontal solid blue line is the white noise power spectrum used for the detector noise simulation.
The three vertical lines correspond to the different channel width for different experiments.
 From left to right, they are 0.1MHz (fiducial, short dash - long dash blue line), 
  8kHz (for MWA, dot - long dash black line), and 4kHz (for LOFAR dot - short dash red line).
}
\end{figure}

The total foreground is so huge that the top
panel looks really similar
to the middle panel (which includes the 21 cm signal). From the
middle panel, we can not really tell if there is any signature of 21cm emission. 
However, the bottom panel shows that the foregrounds can be effectively removed, with the residual 
(recovered 21 cm signal minus the input ``true'' signal) being more than five orders of magnitude below the
foregrounds in amplitude.
% \clearpage
% \clearpage
By transforming the detector noise and the residual from the bottom panel of \fig{midsyn_midff_ps_middet_Fig}
back to $k$-space, we are able to compare them with the 21cm 1D power spectrum, shown in 
\fig{p1d_midsyn_midff_Fig}.
Before foreground cleaning, the total contaminant
(blue dotted curve) is seen to dominate over the 
21cm power spectrum (black solid curve). The foreground power spectrum is seen to rise towards the left, 
reflecting its rather smooth frequency dependence. 
After foreground
cleaning, the residual contaminant (red dashed curve) is significantly below both the original contaminant 
except on scales $k\ll 0.1h/$Mpc.
The flat section of the residual power spectrum for $k\simgt 1h/$Mpc is seen to correspond 
to detector noise. 

The three vertical lines in \fig{p1d_midsyn_midff_Fig} correspond to different minimum channel width for different 
upcoming experiments.
From left to right, they are 0.1MHz (fiducial), 
8kHz (for MWA\footnote{http://web.haystack.mit.edu/MWA/MWA.html})
% (for PAST\footnote{http://astrophysics.phys.cmu.edu/~jbp}) 
and 4kHz 
(for LOFAR\footnote{http://www.lofar.org}). 
Information to the right
of this minimum channel width line is lost, roughly corresponding an exponential
blow-up of the effective detector noise. In other words, the effective detector noise
goes much higher above the signal to the right of those lines so that little
information can be extracted there. 
So in order to 
take advantage of our method of foreground
removal, the channel width needs to be small enough to reach the noise-dominated region. 

% \clearpage
% \clearpage
In other words, 21 cm tomography is limited mainly by foregrounds for $k\ll 1h/$Mpc
and limited mainly by noise for $k\gg 1h/$Mpc.
To take full advantage of their sensitivity by pushing residual foregrounds down to the detector  
noise levels, experiments should therefore be designed to have a 
channel width substantially smaller than 0.1 MHz.
Such small channel widths are realistic and achievable for upcoming experiments.
Since the analysis can now practically be done by dedicated high speed electronics, even
if the software solution was not fast enough. 
% Since the analysis is now all done in software (essentially by fast Fourier transforms), one
% can extract an arbitrarily high channel width as long as there is sufficient 
% CPU power to do it fast enough.

To test the robustness of our foreground cleaning technique, we repeated the above analysis for a wide range 
of foreground models with the same noise level.

First we tested a suite of models with only detector noise and synchrotron radiation, changing the values of the parameters
defined in \eq{synrunningeq}. Most of the results were similar to those shown in
\fig{midsyn_midff_ps_middet_Fig}. 
Increasing the synchrotron amplitude parameter $A_{syn}$
all the way up to completely unrealistic value $10^7\K$ had essentially no effect, because in this case
the simulated spectrum had the exact same shape as the model fit for in the cleaning step.
Likewise, changing the spectral index $\alpha_{syn}$ over the extreme range
$-80 < \alpha_{syn} < 20$ had little effect. 
Complicating the synchrotron spectrum with a running of the running term $\Delta\alpha_{syn}^{rr}\lg ^2({\nu\over\nu_*})$,
so that the intensity of the synchrotron foreground can be written as 
\beq{synrunningeq2}
I_{syn}=A_{syn}\left(\nu\over{\nu_*}\right)^{-\alpha_{syn}-\Delta\alpha_{syn}\lg{{\nu}\over{\nu_*}}-\Delta\alpha_{syn}^{rr}\lg ^2({\nu\over\nu_*})}, 
\eeq
still caused a negligible increase of the residual as long as $|\Delta\alpha_{syn}^{rr}|\le 50$.
This is a reflection of the fact that over a fairly narrow frequency range, 
even a more complicated function can be accurately approximated by a parabola in log-log, \ie, by the
first three terms of its Taylor expansion.

Making variations around the baseline case of multiple foregrounds, 
similarly we tried numerous examples with either higher
foreground amplitudes, different spectral indices or larger running of the spectral indices,
again obtaining results 
similar to the ones shown in \fig{midsyn_midff_ps_middet_Fig}. For example, for the point source foreground, we
tried different values for parameter $\sigma_\alpha$ in \eq{psgaus2eq} from $\sigma_\alpha=0.2$ up to $\sigma_\alpha=100$. We found that
as long as $\sigma_\alpha\le60$ (a conservative cut), 
the residuals are rather insensitive to 
the distribution of point source spectral indices. 

All our tests show that as with astrophysically plausible foreground amplitudes, 
the effectiveness of our simple ``blind'' cleaning method is almost independent
of the number, shape and amplitude of relatively smooth foregrounds.

The basic reason for this robustness is easy to understand. 
As long as the total foreground signal can be well-approximated by our fitting function (a log-log parabola)
over the small frequency interval in question, then the main contribution to the residual will not be the foregrounds,
but the amplitude of this log-log parabola contributed by random detector noise.
Our numerical examples show that the residual indeed does have roughly the shape of our fitting function, not
the shape of the main leading order contribution from residual foregrounds (the next term in their Taylor expansion,
\ie, a cubic term).

% \subsection{Baseline example 2 --- reality (noise$\gg$signal)}
\subsection{Baseline example 2 --- near term situation (noise$\gg$signal)}

The detector white noise we assumed in previous example is small comparing to the signal, in which case we
can subtract the foreground easily for each pixel. Nevertheless, that level of noise might not be achieved
until future next generation experiments. For upcoming experiments, as we showed in \eq{lofarnoise}
and \eq{mwanoise}, the detector noise is far above the 21cm signal. 

In this section, we study the close term 
case by assuming detector noise $\sigma=200\mK$, 200 times larger than previous example, 
 using the same baseline foreground model as in previous section. 

One would expect that for each individual pixel, the huge white noise 
destroys much of the information about the 21 cm signal and
also obscures the frequency dependence of the foregrounds, makeing them harder to fit and remove,
and that the residual will be noise-dominated. 
In this scenario, single pixel cleaning is not enough for cleaning purposes and multiple pixels are needed to average
down the noise and fit the foregrounds. 
However, complications arise when processing multiple pixels. Different pixels come from
different lines of sight, so their 21cm signals are either slightly different realizations or
completely independent realizations of \eq{FT1deq}, depending on how far apart the pixels are
from one another. Furthermore, for pixels that are close to one another, they have slightly
different signals in general, but on large scales, the signals within these pixels are more or less identical,
while on small scales, the signals are almost independent of signals in other pixels. 
The detailes of this complication would probably best be treated via a detailed 3D numerical simulation,
where thousands of pixels can be simulated and the signals from them tested. Since this is beyond
the scope of this paper, we will simply illustrate the basic effects by two extreme
situations, which can also be applied to numerically generated signals. 
% Nevertheless as we will see in following two examples, in principal, our
We will see from the following examples that our 
method for foreground cleaning still works reasonably well.
% This is due to the fact that our
%strategy is aiming at effective foreground removal, and the foreground cleaned
%signal is in good agreement with the simulated 21cm signal plus noise, 
%which is robust under different scenarios.  

\subsubsection{Coherent signal approximation}

For closely separated pixels, we make the crude assumption that the line of sight 21cm signal in these
pixels are identical (same phase and amplitude as in \eq{FT1deq}). 
This approximation will simplify our
calculation, yet the procedure of doing foreground removal is similar 
to generally incoherent signals as we discuss in the next subsection.  

Since the signals are coherent, the total signal for different pixels is the summation of the 
same signal and different foregrounds and noise. We use the same method as described in
\subsec{lownoiseresult} to remove the foreground from total signal along line of sight for each
individual pixel. We then
average all of them in real space and obtain the averaged cleaned signal. 
% \clearpage
\begin{figure}[tb] 
%\vskip-1.0cm
\centerline{\epsfxsize=8.5cm\epsffile{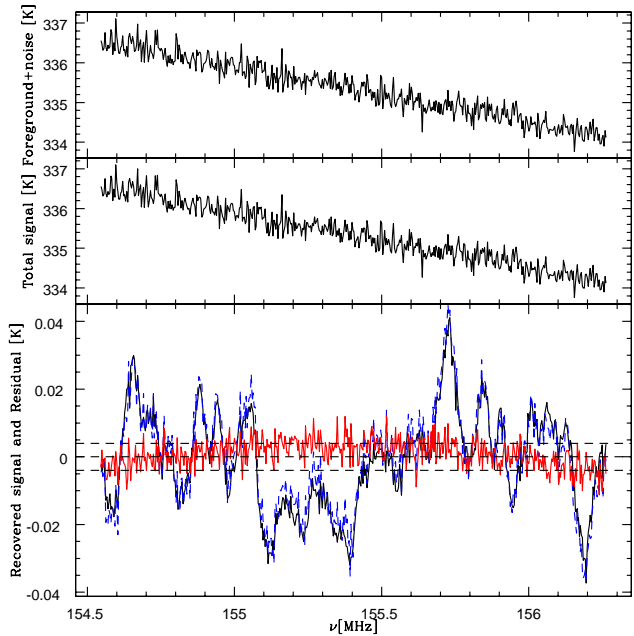}}
% \centerline{\includegraphics[width=8.5cm]{f5.ps}}
\bigskip
%\vskip-1.0cm
\caption{\label{midsyn_midff_ps_highdet_Fig}%
Same as \fig{midsyn_midff_ps_middet_Fig}, but with 200 times larger detector noise ($\sigma=200\mK$).
In the bottom panel, the recovered 21cm signal (blue curve) and the residual (red curve) is the
result of averaging 4000 pixels. All other curves in the plot are from a single pixel. 
}
\end{figure}
% \clearpage
\Fig{midsyn_midff_ps_highdet_Fig} shows the results before and after cleaning. 
The top and middle panels are plotted for a single pixel with 200mK noise. 
The noise wiggles fast on top of the foreground and dominates the signal. It is impossible to tell
the difference between situations with and without 21cm signal (top and middle panels).  
The bottom panel shows the cleaned signal and residual by applying our method and combining 4000
such pixels.
Although both foregrounds and noise are at a level orders of magnitude higher than the signal, the resulting
cleaned signal still captures the main features of the ``true" signal and the residual is well
controlled. 

This confirms that the foregrounds can still be removed effectively when the noise is orders of magnitude
higher than the signal. Our fitting method does not introduce additional contamination to the signal, {\it even when 
foregrounds with many different spectral shapes are averaged together}.

% \clearpage
\begin{figure}[tb] 
%\vskip-1.0cm
\centerline{\epsfxsize=8.5cm\epsffile{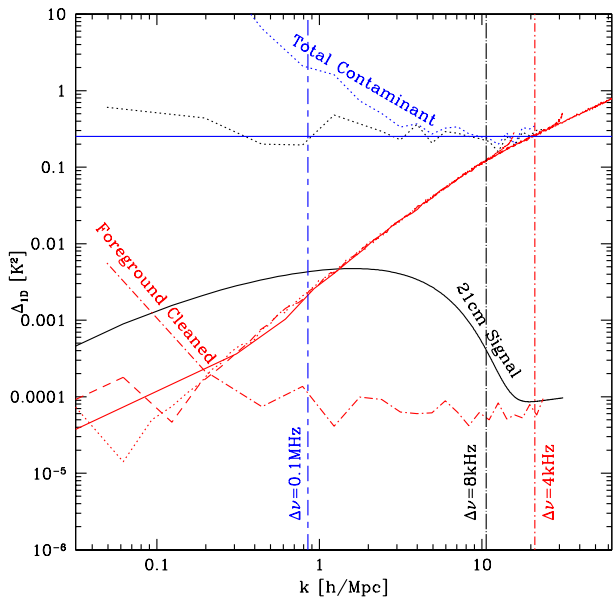}}
% \centerline{\includegraphics[width=8.5cm]{f6.ps}}
% \centerline{\epsfxsize=8.5cm\epsffile{fft_fxnu_det_1em3_syn_a4e7_ff_a3e5_pps_s01_9_b.ps}}
\bigskip
%\vskip-1.0cm
\caption{\label{p1d_midsyn_midff_highdet_Fig}%
Same as \fig{p1d_midsyn_midff_Fig}, but with 200 times larger detector noise ($\sigma=200\mK$). The
coherent case residual contaminants (dot-dashed red curve) are FFT from coherently
averaged residual in
\fig{midsyn_midff_ps_highdet_Fig} (red curve). 
The incoherent case residual contaminant (red solid, dashed, and dotted curves) are
computed from incoherently averaging 40000 pixels. 
The red solid, dashed, and dotted curves are assuming
1.7\MHz, 8.6\MHz, and 17.2\MHz\ bandwidth, respectively. 
}
\end{figure}
% \clearpage
\Fig{p1d_midsyn_midff_highdet_Fig} shows $k$-space signal power spectrum compared with foregrounds and noise
power spectra for single pixel and residual power spectrum from averaging 4000 pixels. Before cleaning, for
each individual pixel, the signal is completely buried under huge noise and foregrounds. After cleaning, the
residual (red dot-dashed curve) is orders of magnitude below both signal and original contaminants except on very large scales.
Similar to \fig{p1d_midsyn_midff_Fig}, the plot suggests that we need frequency resolution substantially
smaller than 0.1MHz to take advantage of foreground removal sensitivity. 

The number of pixels needed to average down the noise varies sharply with the actual noise level. To achieve a similar
level of accuracy shown in  \fig{midsyn_midff_ps_highdet_Fig}, for noise $\sim 500\mK$, we need
approximately 20000 to 30000 pixels. For noise $\sim 100\mK$, we need around 1000 pixels. If the noise is
about the same level as the signal $\sigma =20\mK$, we only need 50 to 100 pixels to adequately lower the noise. 

When larger numbers of pixels getting combined, several side effects appear such as the variation of spectral indices
among different pixels, signal and foreground angular correlation, overestimation of
the sensitivity at small scales, \etc\ 
We will discuss some of these issues a bit more in the next subsection. 
The best approach, however, is to
combine this method with angular approach and remove the contaminants 
in 3D \citep{McQuinn05,MoralesBowmanHewitt05}. Although this is beyond the scope of
the present paper, we hope to address this further in future work.

\subsubsection{Incoherent signals}

For pixels that are far apart, the line of sight 21cm signal in these pixels are no longer the
same. They have different phases and amplitudes and therefore are more or less independent to one another. 
Here we study the case when all signals are completely independent. Compared to the case studied in the previous 
subsection, this case is the opposite extreme.
%We show these two cases for illustrative purposes. 
Signals obtained from real observations will probably have a behavior intermediate between these tow extremes.
Simulating such signals numerically, our cleaning technique is likely to give  residual contaminant levels
that are intermediate between those we find here for these two extreme cases.

When the signals are incoherent, we still use the same method to remove foregrounds from the total signal 
for each individual pixel. However, instead of averaging them in real space, we FFT the signal in each
pixel to Fourier space to obtained the cleaned power spectrum for each pixel. Then we 
average all individual power from different pixels and get the final average cleaned power spectrum. 

% \clearpage
% \begin{figure}[tb] 
% %\vskip-1.0cm
% % \centerline{\epsfxsize=8.5cm\epsffile{p1d_syn335ff33ps01det02_4000.ps}}
% \centerline{\includegraphics[width=8.5cm]{f7.ps}}
% % \centerline{\epsfxsize=8.5cm\epsffile{fft_fxnu_det_1em3_syn_a4e7_ff_a3e5_pps_s01_9_b.ps}}
% \bigskip
% %\vskip-1.0cm
% \caption{\label{p1d_midsyn_midff_incoh_Fig}%
% Same as \fig{p1d_midsyn_midff_highdet_Fig}, but the residual contaminants (red curves) are computed
% from incoherently averaging 40000 signals. The red solid, dashed, and dotted curves are assuming
% 1.7\MHz, 8.6\MHz, and 17.2\MHz\ bandwidth, respectively. 
% }
% \end{figure}
% \clearpage
Comparing with averaging coherent signals in real space, averaging incoherent signals in Fourier space requires larger numbers
of pixels to remove the foregrounds effectively. 
\Fig{p1d_midsyn_midff_highdet_Fig} shows true 21cm power spectrum compared with residual foregrounds and noise
power spectra (red solid, dashed, and dotted curves), defined as the difference between average cleaned power spectrum and true 21cm
power spectrum, from incoherently averaging 40000 pixels. The noise and foregrounds levels are kept the
same as in previous coherent example. Although we average 10 times more
pixels here than for coherent case, the residual contaminant is at a level higher than the
previous case. However, the residual is still reasonable. On most of the scales, the residual contaminant is less than
10\% of the signal. 
Also notice in this case the 21cm power spectrum is best measured for scales around $k=0.1h/\Mpc$, another
consequence from incoherence averaging. In previous two examples, namely low noise and high
noise coherence signal examples, we recovered signal instead of power spectrum.

The three different residual curves in the plot are computed assuming 1.7MHz (red solid curve),
8.6MHz (red dashed curve), and 17.2MHz (red dotted curve) bandwidth, respectively. (We used 1.7MHz
bandwidth for all previous calculations and figures.) As the bandwidth increases, the residual power
decreased especially on smaller $k$. That is to say a larger bandwidth will help foreground
removal at large scales. 

So the bottom line is, for incoherent signals, our method for foreground cleaning still works, yet its
efficiency is reduced due to the fact that signals are independent. In this case, we could consider
increase bandwidth, combine with angular direction measurements, \etc, to improve efficiency and
remove foregrounds effectively.

\section{Discussion}
%\section{Conclusions}
\label{discsec}

We have explored how well foreground contamination can be removed from a 21 cm tomography data cube
by using frequency dependence alone. 
We found that with realistic experimental sensitivities,
21 cm tomography is limited mainly by foregrounds for scales $k\ll 1h/$Mpc
and limited mainly by noise for $k\gg 1h/$Mpc, a result which 
is rather robust to changing the foreground assumptions.
In optimizing the design of upcoming experiments, a
useful rule of thumb is therefore to make the 
channel width substantially smaller than 0.1 MHz, allowing one to 
take full advantage of the detector sensitivity by pushing residual foregrounds down to the 
noise level.
Fortunately, attaining such narrow channel width is realistic for upcoming experiments,
% where the analysis is all done in software (essentially by fast Fourier transforms) and therefore limited 
% only by CPU power.
where the analysis is all done in dedicated high speed electronics.

We used a simple ``blind'' removal technique using no prior information about the nature of the foregrounds, merely 
fitting out a quadratic polynomial in log-log for the frequency dependence separately for each pixel in the sky.
The basic reason that this works so well is that the foregrounds have a much smoother frequency
spectra than the 21 signal. 

Although highly effective, this frequency-based cleaning should be viewed as merely one of three complementary foreground countermeasures.
First, bright point sources can be identified as strong positive outliers, and the corresponding sky pixels can be discarded since
they constitute only a small fraction of the total survey area. 
Second, after the frequency-based cleaning step, noise and signal can be further distinguished by their different angular correlations, 
as described in  Santos {\etal} \citep{Santos04}. This angular approach will be particularly helpful for early 21 cm experiments where
the signal-to-noise ratio is limited.
The angular and frequency-based approaches 
are therefore complementary, and the combination of the two will give the best cleaned 21cm signal with which to study the ``dark'' epoch of
reionization.

Although our results are quite encouraging for the prospects of doing cosmology with 21 cm tomography, 
much work remains to be done on the foreground problem, and we conclude by mentioning a few examples.

A key assumption is this paper is that the foregrounds are dominated by emission mechanisms producing fairly smooth spectra.
The basis for this assumption is that typical atomic and molecular transitions that can produce spectral lines 
correspond to much higher frequencies than those relevant to 21 cm tomography. 
One loophole that needs to checked quantitatively is the possible contribution of recombination lines from hydrogen 
cascading down though very large energy quantum numbers $n\sim 10^7$, although early estimates suggest that this is 
not a significant contaminant \citep[\eg,][]{OhMack03}. 
 
We performed our analysis on simulated data over a small redshift range, limited by our linearization approximation.
It is clearly worthwhile repeating our analysis with a proper hydrodynamical simulation of the 21 cm signal over the full relevant
redshift range. In this case, our 3-parameter foreground fit should be generalized to one that assumes that the
foregrounds are simple only locally in log-frequency space.
An obvious generalization of our method would be to increase the order of the log-log polynomial beyond two.
However, we effectively want to high-pass filter the observed frequency spectrum to clean out foregrounds, and high-order polynominals
can in principle spoil this by having sharp localized features.
A better generalization of our method to long frequency baselines may therefore be either 
a cubic spline in log-log or a Fourier series expansion.
Such end-to-end simulations will also be a valuable tool for quantifying how 
redshift space distortions (whereby the peculiar velocity of the gas breaks the one-to-one correspondence between redshift and frequency)
can be exploited to separate the effects of the matter power spectrum from the ``gastrophysics'' \citep{BarkanaLoeb04}.
This becomes important especially for channel width 
%$\simlt 0.06\MHz$ 
$\simlt 0.1\MHz$ 
\citep{DesjacquesNusser04,Iliev02b}.

In summary, the potential science return from 21 cm tomography is enormous, both for understanding the reionization 
epoch and for probing inflation and dark matter with precision measurements of the small-scale power spectrum.
Our calculations strengthen the conclusion that foreground contamination will not be a show-stopper.
The current situation is similar to the quest for the cosmic microwave background in the 1980's in that the 
cosmological signal has not yet been detected, but better in the sense that the amplitude of both signals and foregrounds
are approximately known, guaranteeing success 
if the engineering challenges can be overcome.

\acknowledgments

The authors wish to thank Angelica de Oliveira-Costa, David Rusin, Gary Bernstein, Jose Diego, 
Lincoln Greenhill, Jacqueline Hewitt, Wayne Hu,
Avi Loeb, Marijke Haverkorn, Miguel Morales and Matias Zaldarriaga 
for helpful discussions and comments.
The authors are especially grateful to Miguel Morales for very informative explanation and discussion on
experimental issues. 
The authors wish to thank the anonymous referee for invaluable comments that improve the paper
substantially. 
XW was supported in part by 
the Kavli Institute for Cosmological Physics through the grant NSF PHY-0114422.   
XW and MT were supported by NASA grant NAG5-11099,
NSF CAREER grant AST-0134999, and fellowships from the David and Lucile
Packard Foundation and the Research Corporation. 
LK and MGS were supported by NSF grant 0307961 and NASA grant NAG5-11098.
MGS was also supported in part by FCT-Portugal under grant BPD/17068/2004/Y6F6.

% \clearpage

%%%%%%%%%%%%%%%%%%%%%%%%%%%%%%%%%%%%%%%%%%%%%%%%%%%%%%%%%%%%
%%%%%%%%%%%%%%%%%%%%%% REFERENCES: %%%%%%%%%%%%%%%%%%%%%%%%%

%\clearpage
%\end{multicols}

%\vskip-1.0cm


\begin{thebibliography}{}

\def\etal{{\it et al.}}

\bibitem[Barkana \& Loeb(2004a)]{BarkanaLoeb04a}
\rf\nn Barkana R\dualand\nn Loeb A;2005;ApJ;626;1
%  astro-ph/0410129
%   Title: Detecting the Earliest Galaxies Through Two New Sources of 21cm Fluctuations
%   Authors: Rennan Barkana (Tel Aviv U.), Abraham Loeb (Harvard U.)
% Journal-ref: Astrophys.J. 626 (2005) 1-11


\bibitem[Barkana \& Loeb(2004b)]{BarkanaLoeb04}
\rf\nn Barkana R\dualand\nn Loeb A;2005;ApJ;624;L65
% astro-ph/0409572
% Title: Line-of-Sight Anisotropy of the 21cm Fluctuations Prior to Reionization
% Authors: Rennan Barkana (Tel Aviv U.), Abraham Loeb (Harvard U.)
% Comments: 4 pages, 2 figures, Phys. Rev. Lett., submitted
% Journal-ref: Astrophys.J. 624 (2005) L65-L68


\bibitem[Becker {\etal}(2001)]{Becker01}
\rf\nnn Becker R H {\etal};2001;AJ;122;2850
% astro-ph/0108097 [abs, ps, pdf, other] :
% Title: Evidence for Reionization at z ~ 6: Detection of a Gunn-Peterson Trough in a z=6.28 Quasar
% Authors: Robert H. Becker, Xiaohui Fan, Richard L. White, Michael A. Strauss, Vijay K. Narayanan, Robert H. Lupton, James E. Gunn, James Annis, Neta A. Bahcall, J. Brinkmann, A. J. Connolly, Istvan Csabai, Paul C. Czarapata, Mamoru Doi, Timothy M. Heckman, G. S. Hennessy, Zeljko Ivezic, G. R. Knapp, Don Q. Lamb, Timothy A. McKay, Jeffrey A. Munn, Thomas Nash, Robert Nichol, Jeffrey R. Pier, Gordon T. Richards, Donald P. Schneider, Chris Stoughton, Alexander S. Szalay, Anirudda R. Thakar, D. G. York
% Comments: Revised version (2001 Sep 4) accepted by the Astronomical Journal (minor changes)
% Journal-ref: Astron.J. 122 (2001) 2850

\bibitem[Bennett {\etal}(2003)]{Bennett03}
\rf\nn Bennett C {\etal};2003;ApJ;148;97
% astro-ph/0302208
% Title: First Year Wilkinson Microwave Anisotropy Probe (WMAP) Observations: Foreground Emission
% Authors: C. Bennett (1), R. S. Hill (2), G. Hinshaw (1), M. R. Nolta (3), N. Odegard (2), L. Page (3), D. N. Spergel (3), J. L. Weiland (2), E. L. Wright (4), M. Halpern (5), N. Jarosik (3), A. Kogut (1), M. Limon (1,6), S. S. Meyer (7), G. S. Tucker (1,6,8), E. Wollack (1) ((1) NASA's GSFC, (2) SSAI (3) Princeton, (4) UCLA, (5) UBC (6) NRC (7) U. Chicago (8) Brown)
%  Journal-ref: Astrophys.J.Suppl. 148 (2003) 97

\bibitem[Bowman {\etal}(2005)]{Bowman05}
\rfprep\nnn Bowman J D, \nnn Morales M F\multiand\nnn Hewitt J N;2005;astro-ph/0512262
% Constraints on Fundamental Cosmological Parameters with Upcoming Epoch of Reionization Observations

\bibitem[Carilli {\etal}(2004b)]{Carilli04}
\rf\nn Carilli C, \nn Furlanetto S, \nn Briggs F, \nn Jarvis M, \nn Rawlings S\multiand\nn
Falcke H;2004;New Astron Rev;48;1029
% astro-ph/0409312 
% Title: Probing the Dark Ages with the Square Kilometer Array
% Authors: C. Carilli, S. Furlanetto, F. Briggs, M. Jarvis, S. Rawlings, H. Falcke 
% Comments: to appear in: "Science with the Square Kilometer Array," New Astronomy Reviews, eds. C. Carilli, S. Rawlings (Elsevier: Amsterdam)
% Journal-ref: New Astron.Rev. 48 (2004) 1029-1038


\bibitem[Carilli {\etal}(2004a)]{Carilli04a} 
\rf\nnn Carilli C L, \nn Gnedin N, \nn Furlanetto S\multiand\nn Owen F;2004;New Astron. Rev.;48;1053
% astro-ph/0409311 
%   Title: Observations of HI 21cm absorption by the neutral IGM during the epoch of re-ionization with the Square Kilometer Array
%   Authors: C.L. Carilli (NRAO), N. Gnedin (U. Colorado), S. Furlanetto (Caltech), F. Owen (NRAO)
%   Comments: espcrc2.sty 9 pages. to appear in: "Science with the Square Kilometer Array," eds. C. Carilli and S. Rawlings, New Astronomy Reviews (Elsevier: Amsterdam)
%   Journal-ref: New Astron.Rev. 48 (2004) 1053-1061

\bibitem[Carilli, Gnedin, \& Owen(2002)]{Carilli02}
\rf\nn Carilli C, \nnn Gnedin N Y\multiand\nn Owen F;2002;ApJ;577;22
% astro-ph/0205169 [abs, ps, pdf, other] :
% Title: HI 21cm absorption beyond the epoch of re-ionization
% Authors: C. Carilli (NRAO), N.Y. Gnedin (Colorado), F. Owen (NRAO)
% Journal-ref: Astrophys.J. 577 (2002) 22-30

\bibitem[Ciardi \& Madau(2003)]{CiardiMadau03}
\rf\nn Ciardi B\dualand\nn Madau P;2003;ApJ;596;1
% astro-ph/0303249
% Authors: B. Ciardi (MPA), P. Madau (UCSC, Oaa)
% Title: Probing Beyond the Epoch of Hydrogen Reionization with 21 Centimeter Radiation
% Journal-ref: Astrophys.J. 596 (2003) 1-8

\bibitem[Desjacques \& Nusser(2004)]{DesjacquesNusser04}
\rf\nn Desjacques V\dualand\nn Nusser A;2004;MNRAS;351;1395
% Astro-ph/0401544
% Title: Redshift distortions in one-dimensional power spectra
% Authors: Vincent Desjacques, Adi Nusser
% Journal-ref: Mon.Not.Roy.Astron.Soc. 351 (2004) 1395
 

\bibitem[Dimatteo, Ciardi, \& Miniati(2004)]{DiMatteo04}
\rf\nn {Di Matteo} T, \nn Ciardi B\multiand\nn Miniati F;2004;MNRAS;355;1053
% astro-ph/0402322
% Title: The 21 centimeter emission from the reionization epoch: extended and point source foregrounds
% Authors: Tiziana Di Matteo, Benedetta Ciardi, Francesco Miniati (MPA)
% Journal: Monthly Notices of the Royal Astronomical Society, Volume 355, Issue 4, pp. 1053-1065, 2004

\bibitem[DiMatteo {\etal}(2002)]{DiMatteo02}
\rf\nn {Di Matteo} T, \nn Perna R, \nn Abel T\multiand\nnn Rees M J;2002;ApJ;564;576
% astro-ph/0109241
% Radio Foregrounds for the 21cm Tomography of the Neutral Intergalactic Medium at High Redshifts
% Authors: Tiziana Di Matteo (1), Rosalba Perna (1), Tom Abel (1,2), Martin J. Rees (2) ((1)Harvard, (2) IoA, Cambridge)
% Journal-ref: 2002ApJ...564..576D

\bibitem[Fan {\etal}(2002)]{Fan02}
\rf\nn Fan X \etal;2002;AJ;123;1247
% astro-ph/0111184
% Title: Evolution of the Ionizing Background and the Epoch of Reionization from the Spectra of z~6 Quasars
% Authors: Xiaohui Fan, Vijay K. Narayanan, Michael A. Strauss, Richard L. White, Robert H. Becker, Laura Pentericci, Hans-Walter Rix

\bibitem[Field(1958)]{Field58}
\rf\nnn Field G B;1958;Proc. IRE;46;240

\bibitem[Field(1959)]{Field59}
\rf\nnn Field G B;1959;ApJ;129;551

\bibitem[Furlanetto \& Briggs(2004)]{FurlanettoBriggs04}
\rf\nn Furlanetto S\dualand\nn Briggs F;2004;New Astron. Rev.;48;1039
% astro-ph/0409205 [abs, ps, pdf, other] :
% Title: 21 cm Tomography of the High-Redshift Universe with the Square Kilometer Array
% Authors: Steven Furlanetto (Caltech), Frank Briggs (RSAA and ATNF)
% Comments: to appear in "Science with the Square Kilometer Array," eds. C. Carilli and S. Rawlings, New Astronomy Reviews (Elsevier: Amsterdam), corrected Fig. 7
% Journal-ref: New Astron.Rev. 48 (2004) 1039-1052

\bibitem[Furlanetto, Sokasian, \& Hernquist(2004)]{Furlanetto03}
\rf\nnn Furlanetto S R, \nn Sokasian A\multiand\nn Hernquist L;2004;MNRAS;347;187
% astro-ph/0305065
% Title: Observing the reionization epoch through 21 centimeter radiation
% Journal-ref: Mon.Not.Roy.Astron.Soc. 347 (2004) 187

\bibitem[Furlanetto, Zaldarriaga, \& Hernquist(2004)]{FZH04}
\rf\nn Furlanetto S, \nn Zaldarriaga M\multiand\nn Hernquist L;2004;ApJ;613;16
% astro-ph/0404112 
%   Title: Statistical Probes of Reionization With 21 cm Tomography
%   Authors: Steven Furlanetto (1), Matias Zaldarriaga (2), Lars Hernquist (2) ((1) Caltech, (2) Harvard University)      
% Journal-ref: Astrophys.J. 613 (2004) 16-22


\bibitem[Gnedin \& Shaver(2003)]{GnedinShaver03}
\rf\nnn Gnedin N Y\dualand\nnn Shaver P A;2004;ApJ;608;611
% Astro-ph/0312005
% Title: Redshifted 21cm Emission from the Pre-Reionization Era: I. Mean Signal and Linear Fluctuations
% Authors: Nickolay Y. Gnedin and Peter A. Shaver
% Journal-ref: Astrophys.J. 608 (2004) 611-621

\bibitem[Gunn \& Peterson(1965)]{GunnPeterson}
\rf\nnn Gunn J E\dualand\nnn Peterson B A;1965;ApJ;142;1633

\bibitem[Haiman \& Holder(2003)]{HaimanHolder03}
\rf\nn Haiman Z \dualand\nnn Holder G P;2003;ApJ;595;1
% astro-ph/0302403
% Title: The Reionization History at High Redshifts I: Physical Models and New Constraints from CMB Polarization
% Authors: Zoltan Haiman (Columbia University), Gilbert P. Holder (IAS)
% Journal-ref: Astrophys.J. 595 (2003) 1-12

\bibitem[Haslam {\etal}(1982)]{Haslam82}
\rf\nnnn Haslam C G T, \nn Stoffel H, \nnn Salter C J\multiand\nnn Wilson W E;1982;A\&AS;47;1
% About 408MHz sky map

\bibitem[Haverkorn, Katgert, \& de Bruyn(2003)]{Haverkorn03}
\rf\nn Haverkorn M, \nn Katgert P\multiand\nn {de Bruyn} G;2003;A\&A;403;1031
% astro-ph/0303575
% Title: Multi-frequency polarimetry of the Galactic radio background around 350 MHz: I. A region in Auriga around l = 161, b = 16
% Authors: Marijke Haverkorn (1), Peter Katgert (1), Ger de Bruyn (2) ((1) Leiden, Netherlands, (2) ASTRON, Netherlands)
% Journal-ref: Astron.Astrophys. 403 (2003) 1031-1044

\bibitem[Hewitt(2004)]{Hewitt04}
\rn Hewitt, J. N., private communications.
% About LOFAR USA sensitivity and beam width
% From: "Jacqueline N. Hewitt" <jhewitt@space.mit.edu>

\bibitem[Holder {\etal}(2003)]{HHKK03}
\rf\nn Holder G, \nn Haiman Z, \nn Kaplinghat M\multiand\nn Knox L;2003;ApJ;595;13
% astro-ph/0302404 
% Title:The Reionization History at High Redshifts II: Estimating the Optical Depth to Thomson Scattering from CMB Polarization
% Authors: Gilbert Holder, Zoltan Haiman, Manoj Kaplinghat, Lloyd Knox
% Journal-ref: Astrophys.J. 595 (2003) 13-18

\bibitem[Hui, Stebbins, \& Burles(1999)]{Hui99}
\rf\nn Hui L, \nn Stebbins A\multiand\nn Burles S;1999;ApJ;511;L5
% astro-ph/9807190
% Title: A Geometrical Test of the Cosmological Energy Contents Using the Lyman-alpha Forest
% Authors: Lam Hui, Albert Stebbins (Fermilab), Scott Burles (U. Chicago)
% Comments: 13 pages, 2 ps figures, submitted to ApJL, on the projected power spectrum
% Journal-ref: Astrophys.J. 511 (1999) L5-9
     
\bibitem[Iliev {\etal}(2002a)]{Iliev02a}
\rf\nnn Iliev I T, \nnn Shapiro P R, \nn Ferrara A\multiand\nn Martel H;2002;ApJ;572;L123
% astro-ph/0202410
% Title: On the direct detectability of the cosmic dark ages: 21cm emission from minihalos

\bibitem[Iliev {\etal}(2002b)]{Iliev02b}
\rf\nnn Iliev I T, \nn Scannapieco E, \nn Martel H\multiand\nnn Shapiro P R;2003;MNRAS;341;81
% astro-ph/0209216
% Nonlinear clustering during the cosmic dark ages and its effect on the 21cm background from minihalos
% comment: numerical simulation, might not cite this one.
% Journal-ref: Mon.Not.Roy.Astron.Soc. 341 (2003) 81

\bibitem[Knox(2003)]{Knox03}
\rf\nn Knox L;2003;New Astron. Rev.;47;883
% Title: CMB Signatures of Extended Reionization
% Authors: Lloyd Knox (UC Davis)
% Comments: To be published in the proceedings of "The Cosmic Microwave Background and its Polarization", New Astronomy Reviews, (eds. S. Hanany and K.A. Olive)
% Journal: New Astronomy Reviews, Volume 47, Issue 11-12, p. 883-886. (2003)
	
\bibitem[Kogut {\etal}(2003)]{Kogut03}
\rf\nn Kogut A {\etal};2003;ApJS;148;161
% 1st yr WMAP observations: TE polarization

\bibitem[Liddle \& Lyth(2000)]{LiddleLythbook}
\rfbook\nnn Liddle A R\dualand\nnn Lyth D H;2000;Cosmological Inflation and Large-Scale Structure;Cambridge Univ. Press;Cambridge

\bibitem[Loeb \& Zaldarriaga(2004)]{Loeb03}
\rf\nn Loeb A\dualand\nn Zaldarriaga M;2004;PRL;92;211301
% Astro-ph/0312134
% Title: Measuring the Small-Scale Spectrum of Cosmic Density Fluctuations 
%          Through 21cm Tomography Prior to the Epoch of Structure Formation
% Authors: Abraham Loeb and Matias Zaldarriaga
% Journal-ref: Phys.Rev.Lett. 92 (2004) 211301

\bibitem[Madau, Meiksin, \& Rees(1997)]{MMR}
\rf\nn Madau P, \nn Meiksin A\multiand\nnn Rees M J;1997;ApJ;475;492
% astro-ph/9608010
% Authors: Piero Madau, Avery Meiksin, Martin J. Rees
% Title: 21-cm Tomography of the Intergalactic Medium at High Redshift
% Journal-ref: 1997 ApJ, 475, 492

\bibitem[McQuinn {\etal}(2005)]{McQuinn05}
\rfprep\nn McQuinn M, \nn Zahn O, \nn Zaldarriaga M, \nn Hernquist L\multiand\nnn Furlanetto S
R;2005;astro-ph/0512263
% astro-ph/0512263
% Cosmological Parameter Estimation Using 21 cm Radiation from the Epoch of Reionization
% Authors: Matthew McQuinn, Oliver Zahn, Matias Zaldarriaga, Lars Hernquist, Steven R. Furlanetto
% Comments: 18 pages, 12 figures, submitted to ApJ

\bibitem[Morales(2004)]{Morales04}
\rf\nnn Morales M F;2005;ApJ;619;678
% astro-ph/0406662
% Title: Power spectrum sensitivity and the design of epoch of reionization observatories
% Authors: Miguel F Morales
% Journal-ref: Astrophys.J. 619 (2005) 678-683

\bibitem[Morales, Bowman, \& Hewitt(2005)]{MoralesBowmanHewitt05}
\rfprep\nnn Morales M F, \nnn Bowman J D\multiand\nnn Hewitt J;2005;astro-ph/0510027
% astro-ph/0510027 [abs, ps, pdf, other] :
% Title: Improving Foreground Subtraction in Statistical Observations of 21 cm Emission from the Epoch of Reionization
% Authors: Miguel F. Morales, Judd D. Bowman, Jacqueline N. Hewitt

\bibitem[Morales \& Hewitt(2003)]{MoralesHewitt03}
\rf\nnn Morales M F\dualand\nn Hewitt J;2004;ApJ;615;7
% astro-ph/0312437
% Authors: Miguel F. Morales, Jacqueline Hewitt
% Title: Toward Epoch of Reionization Measurements with Wide-Field Radio Observations
% Comments: Include formalism, relating observation signal with theoretical computational quantity. Address symmetries used to subtract
% foregrounds independently with other foregrounds subtraction methods, and address nonGaussianity.
% Journal: The Astrophysical Journal, Volume 615, Issue 1, pp. 7-18. 2004

\bibitem[Oh \& Mack(2003)]{OhMack03}
\rf\nnn Oh S P\dualand\nnn Mack K J;2003;MNRAS;346;871
% astro-ph/0302099
% Title: Foregrounds for 21cm Observations of Neutral Gas at High Redshift
% Authors: S. Peng Oh, Katherine J. Mack (Caltech)
% Journal-ref: Mon.Not.Roy.Astron.Soc. 346 (2003) 871

\bibitem[Peacock(1999)]{Peacockbook}
\rfbook\nnn Peacock J A;1999;Cosmological Physics;Cambridge Univ. Press;Cambridge

\bibitem[Pen, Wu, \& Peterson(2004)]{Pen04}
\rfprep\nn Pen {U-L}, \nn Wu {X-P}\multiand\nn Peterson J;2004;astro-ph/0404083
% astro-ph/0404083 [abs, ps, pdf, other] :
% Title: Forecast for Epoch-of-Reionization as viewable by the PrimevAl Structure Telescope (PAST)
% Authors: Ue-Li Pen Xiang-Ping Wu Jeff Peterson
% Comments: 27 pages 24 figures
      
\bibitem[Platania {\etal}(2003)]{Platania03}
\rf\nn Platania P, \nn Burigana C, \nn Maino D, \nn Caserini E, \nn Bersanelli M, \nn Cappellini B\multiand\nn Mennella A;2003;A\&A;410;847
% astro-ph/0303031 [abs, ps, pdf, other] :
% Title: Full Sky Study of Diffuse Galactic Emission at Decimeter Wavelengths
% Authors: P. Platania, C. Burigana, D. Maino, E. Caserini, M. Bersanelli, B. Cappellini, A. Mennella
% Comments: 10 pages, 16 jpeg figures, accepted to Astronomy & Astrophysics, Comments and figure added
% Bibliographic Code: 2003A&A...410..847P

\bibitem[Purcell \& Field(1956)]{PF56}
\rf\nnn Purcell E M\dualand\nnn Field G B;1956;ApJ;124;542

\bibitem[Rottgering(2003)]{Rottgering03}
\rf\nn Rottgering H \etal;2003;TSRA Symp.;;69
% astro-ph/0307240
% Authors:   H. Rottgering, A. G. de Bruyn, R. P. Fender, J. Kuijpers, M. P. van Haarlem, M. Johnston-Hollitt, G. K Miley
% Title: LOFAR: A new radio telescope for low frequency radio observations: Science and project status
% Journal: In: Texas in Tuscany. XXI Symposium on Relativistic Astrophysics, Florence, Italy, 9-13 December 2002, 
%         Eds.: R. Bandiera, R. Maiolino, F. Mannucci. 
%        Singapore: World Scientific Publishing, ISBN 981-238-580-0, 2003, p. 69 - 76
% 2003tsra.symp...69R

\bibitem[Santos {\etal}(2003)]{Santos03}
\rf\nnn Santos M G, \nn Cooray A, \nn Haiman Z, \nn Knox L
\multiand\nn Ma {C-P};2003;ApJ;598;756
% astro-ph/0305471
% Title: Small-scale CMB Temperature and Polarization Anisotropies due to Patchy Reionization
% Authors: M. G. Santos, A. Cooray, Z. Haiman, L. Knox, C.-P. Ma 
% Comments:
% Journal-ref: Astrophys.J. 598 (2003) 756-766

\bibitem[Santos, Cooray, \& Knox(2004)]{Santos04}
\rf\nnn Santos M G, \nn Cooray A\multiand\nn Knox L;2005;ApJ;625;575
% astro-ph/0408515
% Title: Multifrequency Analysis of 21 cm fluctuations From the Era of Reionization
% Authors: M. G. Santos, A. Cooray, L. Knox
% Journal-ref: Astrophys.J. 625 (2005) 575-587 

\bibitem[Seljak {\etal}(2004)]{sdsslyaf}
\rf\nn Seljak {\etal};2005;Phys. Rev. D;71;103515 
% astro-ph/0407372 
% Title: Cosmological parameter analysis including SDSS Ly-alpha forest and galaxy bias: constraints on the primordial spectrum of fluctuations, neutrino mass, and dark energy
% Authors: U. Seljak, A. Makarov, P. McDonald, S. Anderson, N. Bahcall, J. Brinkmann, S. Burles, R. Cen, M. Doi, J. Gunn, Z. Ivezic, S. Kent, R. Lupton, J. Munn, R. Nichol, J. Ostriker, D. Schlegel, M. Tegmark, D. Van den Berk, D. Weinberg, D. York
% Comments: 21 pages, 17 figures, submitted to PRD
% Journal-ref: Phys.Rev. D71 (2005) 103515

\bibitem[Shaver {\etal}(1999)]{Shaver99}
\rf\nnn Shaver P A, \nnn Windhorst R A, \nn Madau P\multiand\nnn \mbox{de Bruyn} A G;1999;A\&A;345;380
% astro-ph/9901320
% Title: Can the reionization epoch be detected as a global signature in the cosmic background?

\bibitem[Spergel {\etal}(2003)]{Spergel03}
\rf\nnn Spergel D N {\etal};2003;ApJS;148;175
% 1st yr WMAP observations: determination of cosmological parameters

\bibitem[Spergel {\etal}(2006)]{spergel06}
\rfprep\nnn Spergel D N {\etal};2006;astro-ph/0603449
% Comment: 3-year WMAP parameter paper

\bibitem[Tegmark(1997)]{mapmaking}
\rf\nn Tegmark M;1997;ApJ;480;L87
% How to make maps from cosmic microwave background data without losing information.
% ASTROPHYSICAL JOURNAL, 1997 MAY 10, V480 N2 PT2:L87-L90.
% mapmaking.tex

\bibitem[Tegmark(1998)]{Max98foreground}
\rf\nn Tegmark M;1998;ApJ;502;1
% astro-ph/9712038 
% Title: Removing real-world foregrounds from CMB maps
% Author: Max Tegmark
% Journal-ref: ApJ, 502, 1-6 (1998)
      
\bibitem[Tegmark {\etal}(2000)]{Max99foreground}
\rf\nn Tegmark M, \nnn Eisenstein D J, \nn Hu W\multiand\nn {de Oliveira-Costa} A;2000;ApJ;530;133
% Astro-ph/9905257
% Title: Foregrounds and Forecasts for the Cosmic Microwave Background
% Authors: Max Tegmark, Daniel J. Eisenstein, Wayne Hu and Angelica de Oliveira-Costa
% Journal-ref: Astrophysical Journal 530, 133-165, 2000

\bibitem[Tegmark \& de Oliveira-Costa(1998)]{Max98ps}
\rf\nn Tegmark M\dualand\nn {de Oliveira-Costa} A;1998;ApJ;500;L83
% astro-ph/9802123
% Title: Removing point sources from CMB maps
% Authors: Max Tegmark and Angelica de Oliveira-Costa
% Journal-ref:  Astrophys.J. 500 (1998) L83-L86

\bibitem[Tegmark, de Oliveira-Costa, \& Hamilton(2003)]{TOH03}
\rf\nn Tegmark M, \nn {de Oliveira-Costa} A\multiand\nn Hamilton A;2003;Phys. Rev. D;68;123523
% astro-ph/0302496 
% Title: A high resolution foreground cleaned CMB map from WMAP
% Authors: Max Tegmark (Penn), Angelica de Oliveira-Costa (Penn), Andrew Hamilton (JILA)
% Journal-ref: Phys.Rev. D68 (2003) 123523

\bibitem[Tegmark {\etal}(2004)]{sdsspars}
\rf\nn Tegmark M {\etal};2004;Phys. Rev. D;69;103501
% astro-ph/0310723
% Title: Cosmological parameters from SDSS and WMAP
% Authors: M. Tegmark, M. Strauss, M. Blanton, K. Abazajian, S. Dodelson, H. Sandvik, X. Wang, D. Weinberg, I. Zehavi, N. Bahcall, F. Hoyle, D. Schlegel, R. Scoccimarro, M. Vogeley, A. Berlind, T. Budavari, A. Connolly, D. Eisenstein, D. Finkbeiner, J. Frieman, J. Gunn, L. Hui, B. Jain, D. Johnston, S. Kent, H. Lin, R. Nakajima, R. Nichol, J. Ostriker, A. Pope, R. Scranton, U. Seljak, R. Sheth, A. Stebbins, A. Szalay, I. Szapudi, Y. Xu, 27 others (the SDSS collaboration)
% Comments: Minor revisions to match accepted PRD version. SDSS data and ppt figures available at this http URL
% Journal-ref: Phys.Rev. D69 (2004) 103501

\bibitem[Tozzi {\etal}(2000)]{Tozzi99}
\rf\nn Tozzi P, \nn Madau P, \nn Meiksin A\multiand\nnn Rees M J;2000;ApJ;528;597
% astro-ph/9903139
% Authors: Paolo Tozzi, Piero Madau, Avery Meiksin, Martin J. Rees
% Title: Radio Signatures of HI at High Redshift: Mapping the End of the ``Dark Ages''
% Journal-ref: 2000 ApJ, 528,597

\bibitem[Wouth(1952)]{Wouth52}
\rf\nnn Wouthuysen S A;1952;AJ;57;31 

\bibitem[Wyithe \& Loeb(2004a)]{WyitheLoeb04a}
\rf\nn Wyithe S\dualand\nn Loeb A;2004;ApJ;610;117
% astro-ph/0401554 
%   Title: Redshifted 21cm Signatures Around the Highest Redshift Quasars
%   Authors: Stuart Wyithe, Abraham Loeb
% Journal-ref: Astrophys.J. 610 (2004) 117-127


\bibitem[Wyithe \& Loeb(2004b)]{WyitheLoeb04}  
\rf\nn Wyithe S\dualand\nn Loeb A;2004;Nature;432;194
% astro-ph/0409412
% Title: A Size of ~10 Mpc for the Ionized Bubbles at the End of Cosmic Reionization
% Authors: Stuart Wyithe, Abraham Loeb
% Comments: Accepted for publication in Nature. Press embargo until published
% Journal: Nature, Volume 432, Issue 7014, pp. 194-196 (2004)

\bibitem[Zaldarriaga, Furlanetto, \& Hernquist(2003)]{Matias03}
\rf\nn Zaldarriaga M, \nnn Furlanetto S R\multiand\nn Hernquist L;2004;ApJ;608;622
% astro-ph/0311514
% Authors: Matias Zaldarriaga, Steven R. Furlanetto and Lars Hernquist
% Title: 21 Centimeter Fluctuations From Cosmic Gas at High Redshifts
% Journal-ref: Astrophys.J. 608 (2004) 622-635


      
% \bibitem[LOFAR]{lofar}
% \rn LOFAR website, http://www.lofar.org

% \bibitem{mileura}
% % This reference might not be the most appropriate one for mileura
% \rn http://web.haystack.mit.edu/MWA/MWA.html
% % http://www.atnf.csiro.au/

% \bibitem{past}
% \rn For details on PAST, see http://astrophysics.phys.cmu.edu/~jbp

% \bibitem{ska}
% \rn SKA website, http://www.skatelescope.org


\end{thebibliography}
\end{document}